\documentclass[%
 reprint,
 amsmath,amssymb,
 aps,
]{revtex4-2}

\usepackage{amsmath}
\usepackage{amsfonts}

\usepackage{euscript}
\newcommand{\normal}{\EuScript{N}}
\newcommand{\kernel}{\EuScript{K}}
\newcommand{\predmean}{\EuScript{E}}
\newcommand{\predvar}{\Sigma}
\newcommand{\cutoff}{\Theta}

\usepackage{mathtools}
\DeclarePairedDelimiter\paren{(}{)}
\DeclarePairedDelimiter\angl{\langle}{\rangle}
\DeclarePairedDelimiter\determ{\lvert}{\rvert}

\newcommand{\conj}[1]{\paren*{#1}^\ast}

\usepackage[per-mode=symbol]{siunitx}
\DeclareSIUnit{\calorie}{cal}
\usepackage{chemformula}
\usepackage{graphicx}
\usepackage{enumerate}
\usepackage{import} 
\usepackage{hyperref}
\usepackage{booktabs}
\usepackage{color, colortbl}
\usepackage{caption}
\usepackage{subcaption}

\makeatletter
\renewcommand\@makecaption[2]{%
  \par
  \vskip\abovecaptionskip
  \begingroup
   \small\rmfamily
    \begingroup
     \samepage
     \flushing
     \let\footnote\@footnotemark@gobble
     \@make@capt@title{#1}{#2}\par
    \endgroup
  \endgroup
  \vskip\belowcaptionskip
}
\makeatother

\begin{document}

\preprint{APS/123-QED}

\title{Sparse Gaussian Process Potentials:
Application to Lithium Diffusivity in
Superionic Conducting Solid Electrolytes
}

\author{Amir Hajibabaei}
\author{Chang Woo Myung}
\author{Kwang S. Kim}
\email{kimks@unist.ac.kr}
\affiliation{
Center for Superfunctional Materials, 
Department of Chemistry and Department of Physics, 
Ulsan National Institute of Science and Technology, 
Ulsan 44919, Korea
}

\date{\today}

\begin{abstract}
For machine learning of interatomic potentials
a scalable sparse Gaussian process regression formalism is
introduced with a data-efficient on-the-fly adaptive sampling 
algorithm.
With this approach, the computational cost is effectively
reduced to those of the Bayesian linear regression methods
whilst maintaining the appealing characteristics
of the exact Gaussian process regression.
As a showcase, experimental melting and glass-crystallization
temperatures are reproduced for \ch{Li7P3S11},
\ch{Li} diffusivity is simulated, and an unchartered phase
is revealed with much lower \ch{Li} diffusivity which
should be circumvented.
\end{abstract}

\maketitle

\section{Introduction}

A solid with ionic conductivity similar to liquid electrolytes
has far reaching implications for the energy storage industry.
Recently, first-principles (FP) calculations have been
applied extensively for theoretical studies of ionic diffusion 
in solid electrolytes~\cite{Deng2016ComputationalBatteries}.
But due to the huge computational resources required 
for the electronic structure calculations these 
simulations are often applied only to small systems 
($\sim 100$ atoms) and short intervals of time 
($\sim\SI{100}{\pico\second}$).
Moreover, due to the Arrhenius-type temperature 
dependence of the diffusion events in solids,
they occur too slowly at room temperature to 
be sampled with statistical certainty using 
ab initio molecular dynamics (AIMD)~\cite{He2018}.
Therefore often simulations are carried out at 
elevated temperatures and ionic diffusivity 
at room temperature is approximated by extrapolation.

As an alternative to FP calculations,
it has been demonstrated that machine learning (ML)
methods can be applied for accurate representations 
of the potential energy surface (PES).
Most notably, 
neural-network (NN) representations~\cite{Behler2007},
Gaussian approximation potentials (GAP)~\cite{Bartok2010},
gradient-domain~\cite{Chmiela2017} and
symmetrized gradient-domain ML~\cite{Chmiela2018},
deep potential~\cite{Zhang2018},
compressed sensing~\cite{Seko2015First-principlesSensing},
and deep neural-network for molecular wavefunctions~\cite{Schutt2019}
are among various ML techniques 
which have been implemented.
Using these techniques, the PES is learned from the data
(potential energy and forces)
obtained from limited FP calculations.
For generation of the relevant data, on-the-fly 
learning~\cite{Li2015MolecularForces, Jinnouchi2019On-the-flyPoints, Jinnouchi2019PhaseInference, Vandermause2020On-the-FlyEvents},
a query-by-committee active-learning~\cite{Artrith2012High-dimensionalCopper},
and global exploration of PES~\cite{Bernstein2019DeSurfaces}
have been reported.

Training of neural-networks scale linearly with the size of data $n$
but they contain a huge number of optimizable parameters
and require big data to avoid over-fitting.
On the other hand, the kernel methods where a covariance
matrix is constructed and inverted, are (almost) non-parametric 
but scale as $n^2$ for construction and $n^3$ for inversion
of the covariance matrix.
Considering that a single FP calculation for a system with 
$N$ atoms yields $3N+1$ data components (energy and forces), 
the size of covariance matrix becomes huge even with a small 
number of training data.
Because of this unfavorable scaling,
a critical issue is balancing of the global exploration
versus local exploitation in sampling the data.
For instance, in some on-the-fly learning algorithms
the ML model is continuously updated during MD 
which keeps the model locally accurate but 
undermines its transferability.

Here we explore low-rank approximations of the covariance
matrix and sparse Gaussian process regression 
(SGPR)~\cite{Rasmussen2005GaussianLearning, pmlr-v5-titsias09a}
for improving the scalability and an adaptive 
sampling algorithm for optimal on-the-fly generation
of the training data and global exploration.
We study the lithium diffusivity in solid electrolytes 
to explore possible superionic conducting materials
towards fast charging inflammable batteries.

\section{Sparse Gaussian process regression potentials}

Assuming that the force on atom $i$ in 
a chemical system depends only on the relative 
coordinates of atoms $j$ in its neighborhood, 
the local chemical environment (LCE) of $i$ 
is defined as
\begin{equation}\label{eq:lce}
    \rho_i = \{\vec{r}_{ij}; 
               j\ne i \: \& \: 
               |\vec{r}_{ij}|<r_c
             \}
\end{equation}
where $r_c$ is the cutoff for neighborhood relations.
The optimal cutoff can be obtained by balancing the 
errors of the regression algorithm and the random noise 
caused by ignoring the atoms beyond the 
cutoff~\cite{Deringer2017MachineCarbon}.
Since any configuration $x$ of $N$ atoms is compiled to
a list of LCEs ${x}=\{\rho_i\}_{i=1}^{N}$,
the potential energy becomes additive over local terms
\begin{equation}
    E(x) = \sum_{i=1}^N \varepsilon(\rho_i)
\end{equation}
where $\varepsilon$ is a fictional latent function.
The common theme for all Bayesian regression methods
is defining a covariance kernel $\kernel(\rho_{i},\rho_{i'})$
which encodes the similarity of LCEs and is invariant 
with respect to the symmetry operations that leave the 
potential energy invariant 
e.g. translations and rotations~\cite{Bartok2013}.
It models the covariance between local energies:
$\kernel(\rho_{i},\rho_{i'}) \propto \angl*{\varepsilon(\rho_{i})\varepsilon(\rho_{i'})}$.
We assume that the kernel is normalized: 
$0\leq\kernel(\rho_{i},\rho_{i'})\leq 1$.
The kernel for a pair of configurations $(x,x')$ becomes
\begin{equation}
    \kernel(x,x') = \sum_{i\in x, i'\in x'} 
                        \kernel(\rho_{i}, \rho_{i'})
    \texttt{.}
\end{equation}
The training data consists of a set of configurations
${X}=\{{x}_j\}_{j=1}^{n}$ as inputs along with their
potential energies and forces obtained from FP 
calculations as targets $Y$.
For briefness, in the following we only consider 
potential energies $Y=\{E_j\}_{j=1}^{n}$.
By representing $\varepsilon$ as a 
Gaussian process it is assumed that the targets 
are sampled from a distribution given 
by~\cite{Rasmussen2005GaussianLearning}
\begin{equation}
    Y|X \sim \normal 
    \paren*{0, \beta\paren*{k_{nn}+\sigma^2 \mathbb{I}}}
\end{equation}
where $k_{nn}$ is the covariance matrix $\kernel(x_i, x_j)$ 
amongst the data, $\normal$ is a multivariate Gaussian 
distribution, $\beta$ is the scale of variance
and $\sigma$ the scale of random noise in $Y$.
$\sigma,\beta$ and other hyper-parameters which may be
defined in $\kernel$ are optimized by maximizing the
log-likelihood of $Y$.
For a test configuration ${x}^\ast$, the predictive 
distribution $E^\ast$ for the potential energy 
becomes
\begin{gather}\label{eq:gpr}
    E^\ast | x^\ast, X, Y \sim 
    \normal(\predmean, \predvar) 
    \\
    \predmean = k_{\ast n} 
    \paren*{k_{nn} + \sigma^2 \mathbb{I}}^{-1} Y 
    \label{eq:predmean}
    \\
    \predvar = \beta \paren*{
    k_{\ast\ast} -  k_{\ast n} 
    \paren*{k_{nn} + \sigma^2 \mathbb{I}}^{-1}
    k_{\ast n}^\mathsf{T}
    }
\end{gather}
where $k_{\ast n}$ is the covariance matrix for 
${x}^\ast$ and data ${X}$, $\predmean$ is the predictive mean,
and $\predvar$ is the predictive variance.
\autoref{eq:predmean} is the defining equation for
Gaussian process regression (GPR).
This calculation requires inversion of a large covariance
matrix which scales as $\mathcal{O}(n^3)$.

For improving the scalability, one can exploit the 
LCE similarities in the data $X$.
It is very likely that many LCE pairs in $X$
are similar: $\kernel(\rho,\rho')\approx 1$.
Let ${z}=\{\chi_j\}_{j=1}^{m}$ denote a reduced set 
of LCEs which are significantly distinct from each other 
and are sufficient statistics for ${X}$.
$z$ is called the inducing set of LCEs.
The key for a SGPR model 
is low-rank approximation of the covariance 
matrices~\cite{Rasmussen2005GaussianLearning}
\begin{equation}\label{eq:lr}
    \begin{split}
        k_{nn} \approx k_{nm} k_{mm}^{-1} k_{nm}^{\mathsf{T}}
        \\
        k_{\ast n} \approx k_{\ast m} k_{mm}^{-1} k_{nm}^{\mathsf{T}}
    \end{split}
\end{equation}
where $k_{mm}$ is the covariance matrix 
amongst ${z}$, $k_{nm}$ cross covariance 
between $X$ and $z$, and $k_{\ast m}$ 
between $x^\ast$ and $z$.
Using the Woodbury matrix identity,
this results in a predictive mean given by
\begin{gather}
    \predmean \approx k_{\ast m} \mu 
    \label{eq:reg} \\
    \mu = (\sigma^{2} k_{mm} + k_{nm}^\mathsf{T} k_{nm})^{-1}
    k_{nm}^\mathsf{T} Y 
    \label{eq:reg_mu}
    \texttt{.}
\end{gather}
It is worth noting that if 
$k_{mm}$ in the above expression 
is replaced by $\gamma^{-2}\mathbb{I}$, this formalism
will be reduced to the Bayesian linear 
regression (BLR) method~\cite{Jinnouchi2019PhaseInference}.
In linear regression algorithms~\cite{Seko2015First-principlesSensing, Jinnouchi2019PhaseInference},
regardless of a regularization term,
the linear system $k_{nm}\mu=Y$ is solved 
while $\mu$ in \autoref{eq:reg_mu} is the solution of
\begin{equation}
    \begin{bmatrix}
    k_{nm} \\
    \sigma L^\mathsf{T}
    \end{bmatrix}
    \mu 
    =
    \begin{bmatrix}
    Y \\
    \mathbf{0}
    \end{bmatrix}
\end{equation}
where $L$ is the Cholesky factor of $K_{mm}$ 
and $\mathbf{0}$ is a columnar vector of zeros
with length $m$.
Therefore converting BLR to SGPR is straightforward.
On the other hand, normal GPR is recovered if we set $z=X$.
Also note that if we substitute
$\mu=k_{mm}^{-1} w$
in \autoref{eq:reg} it looks like a GPR based on
$({z}, w)$ on behalf of $({X}, Y)$.
Therefore this approximation is also called 
projected process (PP).
SGPR and PP are distinguished only in defining the 
loss function for optimizing the hyper-parameters 
and selection of ${z}$.
In SGPR the variational lower-bound 
of the true log marginal likelihood 
introduced by Titsias~\cite{pmlr-v5-titsias09a}
should be maximized.
Inclusion of the forces data in regression
is formulated in the Supplemental Material 
(SM)~\cite{prl_sm}.
A useful quantity that indicates
the covariance loss due to sparsification is 
\begin{equation} \label{eq:spilling}
    s(\rho) = 
    k(\rho, \rho) -
    k_{\rho m}k_{mm}^{-1}k_{\rho m}^\mathsf{T}
\end{equation}
which is referred to as the spilling factor~\cite{Miwa2016MolecularPotential} 
but it is also proportional to the predictive 
variance of a GPR based on $({z}, w)$.

\begin{figure}
    \centering
    \includegraphics[scale=0.45]{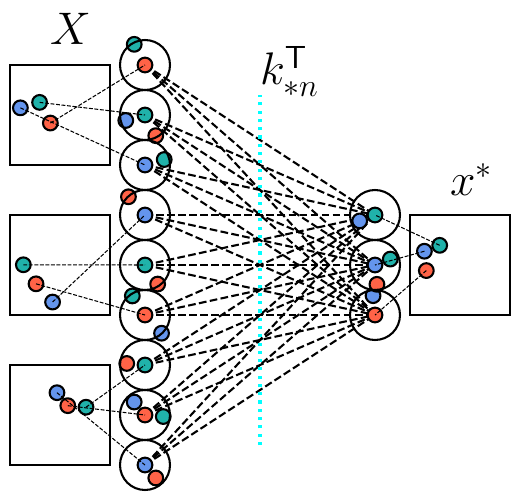}
    \includegraphics[scale=0.45]{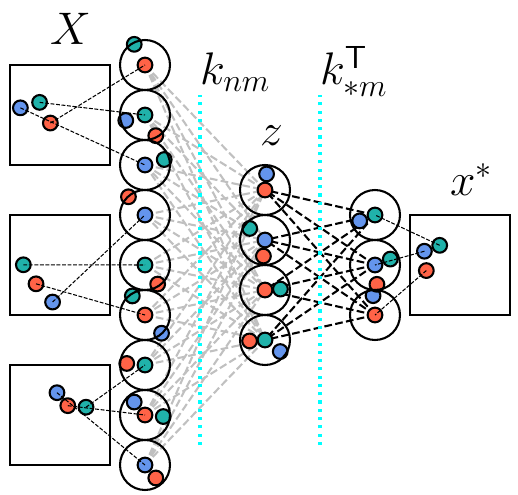}
    \caption{Schematic comparison of GPR (left) and SGPR (right).
    Squares represent atomic configurations.
    Circles represent compilation of configurations into LCEs
    around each atom.
    In GPR the potential energy for $x^\ast$ is predicted
    by calculating its covariance matrix with the data $X$.
    In SGPR only the covariance matrix with the inducing LCEs 
    $z$ is needed.
    }
    \label{fig:schematic}
\end{figure}

After projection, only ${z}$ and $\mu$  
are needed for predictions, reducing
the computational cost to $\mathcal{O}(m)$
instead of $\mathcal{O}(nN)$ (see \autoref{fig:schematic}).
For calculating $\mu$ we only need to invert 
a matrix of size $m$ which is much 
smaller than $(3N+1)n$ as in full GPR.
The total cost for calculation of $\mu$ is proportional to
$(3N+1)n m^2 \ll (3N+1)^3n^3$~\cite{Foster2009StableSatyavolu}.
Considering that the kernel itself is usually 
computationally expensive, the cost of constructing
the related covariance matrices is even more important.
Since in SGPR we only need to calculate the kernel
between the data and inducing LCEs, a significant
improvement is achieved in comparison with full GPR.
In a naive comparison the improvement is
$\mathcal{O}(n^2)\to \mathcal{O}(n)$, 
but a more rigorous analyses 
in SM 
shows a much better improvement.
For example with $100$ configuration ($100$ atoms each), 
the improvement is a factor of $\mathcal{O}(10^3)$.

\section{The kernel}

For a similarity kernel between LCEs,
the smooth overlap of atomic positions (SOAP) 
is defined as~\cite{Bartok2013}
\begin{gather}
    \kernel(\rho_{i},\rho_{i'}) = \int d\mathbf{R}  
    \determ*{
    \int d\vec{r} 
    \xi_{i}(\vec{r}) \xi_{i'}(\mathbf{R} \vec{r}) 
    }^2 
\end{gather}
where $\mathbf{R} $ is the 3D rotations operator,
$\xi_{i}(\vec{r})$ is the atomic density
in neighborhood of $i$
\begin{equation}
    \xi_{i}(\vec{r}) = \sum_{j\in\rho_{i}} 
    e^{ -\alpha |\vec{r}-\vec{r}_{ij}|^2 } 
    \texttt{,}
\end{equation}
and $\alpha$ is a hyper-parameter.
Analytical integration results in~\cite{Bartok2013}
\begin{equation}
    \kernel(\rho_{i},\rho_{i'}) \propto 
    \sum_{l,m,m'} \frac{1}{2l+1} \conj{I^l_{mm'}} I^l_{mm'}
\end{equation}
where
\begin{equation}
    I^l_{mm'} = \sum_{\substack{j\in\rho_{i}\\j'\in\rho_{i'}}}
    e^{ -\alpha (r_{j}^{2} + r_{j'}^{2})/2 }
    \iota_{l}(\alpha r_{j} r_{j'}) 
    {Y_{l}^{m}}(\hat{n}_{j})
    \conj{Y_{l}^{m'}(\hat{n}_{j'})}
    \texttt{,}
\end{equation}
$r_j=|\vec{r}_{ij}|$, $\hat{n}_j = \vec{r}_{ij}/r_j$, 
and $\iota_l$ is the modified Bessel 
function of the first kind (Bessel-i).
$\iota_l$ can be expanded as~\cite{Watson1995AFunctions}
\begin{gather}
    \iota_{l}( \alpha r_j r_{j'} ) = 
    \sum_{n=0}^{\infty} a_{nl} r_j^{2n+l}r_{j'}^{2n+l} 
\end{gather}
where $a_{nl} = {({\alpha}/{2})^{2n+l}}/{(n+l)!n!}$.
Here, we use this expansion to simplify the kernel
into an inner product
\begin{gather}
    c_{nlm}^{i} = \sqrt{a_{nl}}
    \sum_{j\in\rho_{i}}
    \cutoff(r_j)
    e^{ -\alpha r_{j}^{2} /2 }
    r_{j}^{2n+l}
    {Y_{l}^{m}}(\hat{n}_{j})
    \label{eq:cnlm}
    \\
    I^l_{mm'} = \sum_n {c^i_{nlm}} \conj{c^{i'}_{nlm'}}
    \label{eq:ilmm} 
    \\
    p^i_{nn'l} = \sum_{m} \frac{{c^i_{nlm}}
    \conj{c^i_{n'lm}}}{\sqrt{2l+1}} 
    \\
    \kernel(\rho_i,\rho_{i'}) \propto 
    \sum_{n,n',l} p^i_{nn'l} \conj{p^{i'}_{nn'l}}
    \texttt{.}
\end{gather}
$\cutoff$ in \autoref{eq:cnlm} is a smooth cutoff 
function which is manually inserted for elimination 
of discontinuities in the kernel and its gradients 
upon atoms entering to/exiting from the LCEs.
The $a_{nl}$ multipliers decrease extremely fast by
increasing $n$ and $l$; 
therefore it suffices to calculate up to
only a few orders of expansion.
In our calculations $\alpha=1$, $r_c=8$, 
$\cutoff(r)=(1-r/r_c)^2$, and $l,n\leq 3$.

Extension to multi-species environments is 
straightforward.
Species-dependent LCE becomes $\rho_i=\{\rho_i^\alpha\}$
where $\alpha$ indexes the species present in $\rho_i$.
Then the coefficients in \autoref{eq:cnlm} are 
calculated separately for each atomic species 
($c^i_{nlm}\to c^{i\alpha}_{nlm}$) and
the total descriptor vector becomes~\cite{De2016ComparingSpace}
\begin{equation}
    p^{i\alpha\beta}_{nn'l} = \frac{1}{\sqrt{2l+1}} 
     \sum_{m}{c^{i\alpha}_{nlm}} \conj{c^{i\beta}_{n'lm}}
     \texttt{.}
\end{equation}
The generalized multi-species kernel
becomes~\cite{De2016ComparingSpace}
\begin{equation}
    \kernel(\rho_{i}, \rho_{i'}) = \mu_{s_{i}s_{i'}}
    \paren*{ 
    \sum_{\substack{\alpha,\alpha',\beta,\beta', \\ n, n', l}}
    p^{i\alpha\beta}_{nn'l} 
    \conj{p^{i'\alpha'\beta'}_{nn'l}}
    \nu_{\alpha\alpha'}\nu_{\beta\beta'}
    }^\eta
\end{equation}
where $\nu_{\alpha\beta}$ is an alchemical similarity
kernel between the atomic species in LCEs
and $\mu_{s_{i}s_{i'}}$ is a kernel
which depends on the species of the atoms $i,i'$
and quantifies the covariance between local energies.
Here we have chosen $\nu_{\alpha\beta}=\delta_{\alpha\beta}$
and $\mu_{s_{i}s_{i'}}=\delta_{s_{i}s_{i'}}$. 
The latter corresponds to summation of 
independent Gaussian processes defined
for each atomic species.
The exponent $\eta$ controls the sensitivity of the kernel
which is fixed at $4$ in this work.
As usual, the kernel is normalized by 
\begin{equation}
    \kernel(\rho_{i}, \rho_{i'}) \to 
    \frac{
        \kernel(\rho_{i}, \rho_{i'})
        }
        {
        \sqrt{\kernel(\rho_{i}, \rho_{i})
              \kernel(\rho_{i'}, \rho_{i'})
              }
    }
    \texttt{.}
\end{equation}

\section{On-the-fly sampling}

The most important step, regardless of the regression
algorithm, is generation of the data.
Early methods~\cite{Behler2007, Bartok2010} prepared 
the data beforehand which is not efficient.
On-the-fly learning 
methods~\cite{Li2015MolecularForces, Jinnouchi2019On-the-flyPoints, Jinnouchi2019PhaseInference, Vandermause2020On-the-FlyEvents}
estimate the accuracy of the potential 
and use a criterion for data generation.
In GPR based algorithms the predictive variance 
is a suitable criterion while in NNs the uncertainty 
can be emulated by training identical NNs with 
different initial random seeds.
Generally, GPR based algorithms are more suited
for on-the-fly  learning since they have only a 
few hyper-parameters.
But the size of these potentials and their 
computational cost grow with time, therefore 
the data generated earlier need to be 
dumped to keep the potential affordable.
Thus for globally accurate potentials, 
it is critical to devise an optimal 
adaptive sampling algorithm.
We tackle this issue by replacing the 
predictive variance by another criterion
which we call the geometric criterion.
This is motivated from the observation that the predictive 
variance does not take advantage of energy and forces data.
Briefly, aside from hyper-parameter optimization,
a SGPR potential can be modified in two ways: adding a LCE $\chi$
to the inducing set $z$, adding a snapshot $x$ to the data $X$.
Instead of calculating the predictive variance, we calculate the 
change in the predictive mean of the potential by directly 
constructing the potential both with and without $x$ (same for $\chi$).
If the change in the predictive mean is more than a 
threshold $\epsilon$ the addition is accepted
(See SM~\cite{prl_sm} for algorithm).

\section{Computational details}

The \texttt{VASP}~\cite{Kresse1996EfficientSet} package which implements the
projector augmented-wave~\cite{Blochl1994ProjectorMethod}
approach to DFT with
PBE GGA functionals~\cite{Perdew1996RationaleApproximations}
is used for all FP calculations.
Calculations are non-spin-polarized and 
kinetic energy cutoff of \SI{500}{\electronvolt}
is applied.
We have developed the python package 
\href{https://github.com/amirhajibabaei/AutoForce}
{\texttt{AutoForce}} for generating SGPR models
and on-the-fly learning of the PES~\cite{Ha}.
This package is coupled with the atomic simulation
environment (\texttt{ASE})~\cite{HjorthLarsen2017TheAtoms}
and from there it can be coupled to various
FP software including \texttt{VASP}.
Phonon spectra for \ch{Si} is calculated 
using the \texttt{phonopy}~\cite{Togo2015FirstScience} 
package.

MD with a ML potential is referred to as MLMD.
For reporting the accuracy of ML potentials, 
we use root mean squared error (RMSE) 
and coefficient of determination
which is denoted as $R^2$ in statistics
and is defined by
\begin{equation}
    R^2 = 1 - \frac{\sum_i (f_i-\widetilde{f}_i)^2}
    {\sum_i (f_i-\overline{f})^2}
\end{equation}
where $\{f_i\}$ are FP forces, 
$\overline{f}$ is their average,
and $\{\widetilde{f}_i\}$ are ML forces.
$R^2$ indicates the fraction of explained variance.

\section{Silicon}

\begin{figure}
    \centering
    \includegraphics[scale=0.9]{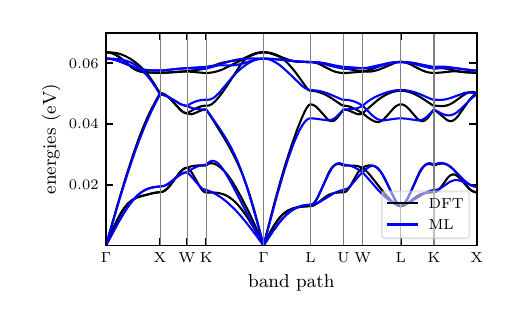}
    \caption{Phonon spectra of bulk Si with a 
    SGPR potential generated using only 13 DFT 
    data and 36 inducing LCEs.}
    \label{fig:si}
\end{figure}

\begin{figure}
    \centering
    \includegraphics[scale=0.9]{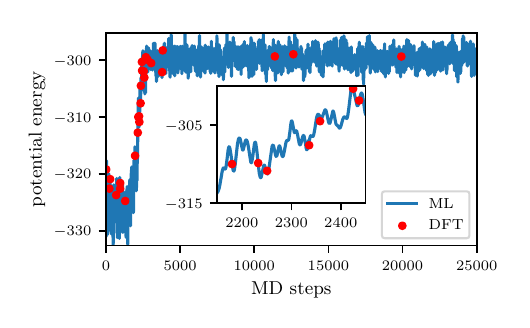}
    \caption{On-the-fly MLMD with the adaptive sampling algorithm
    for \ch{Si} at \SI{2000}{\kelvin}.
    The time-step for MD is \SI{2}{\femto\second}.
    Melting occurs at steps $\sim2000$ to $\sim 3000$.
    The system is liquid beyond that.}
    \label{fig:si_melting}
\end{figure}

For a benchmark simulation, we created a SGPR
potential for bulk \ch{Si} since it has been
studied by several other ML 
methods~\cite{Behler2007, Bartok2010, Li2015MolecularForces, Jinnouchi2019On-the-flyPoints, Vandermause2020On-the-FlyEvents}.
A cubic cell with 64 atoms was chosen with a
$2\times 2\times 2$ k-point grid for DFT calculations.
The potential is generated by on-the-fly sampling
with MD simulations
at \SI{300}{\kelvin} (for \SI{20}{\pico\second})
and \SI{1000}{\kelvin} (for \SI{30}{\pico\second}).
Using the algorithm ULTRAFAST (see SM), 
only $13$ DFT calculation are performed for data
generation and $36$ LCEs are sampled as the 
inducing set. 
This entire simulation took less than \SI{1}{\hour}.
$100$ snapshots are chosen from trajectories
for testing which result in the 
RMSE=\SI{0.07}{\electronvolt\per\angstrom} 
and $R^2=0.994$ for forces.
Additionally we tested the model by calculating
the phonon spectra with a $5\times 5\times 5$
supercell which is shown in \autoref{fig:si}.
Direct comparison with previous methods is not 
possible since they model \ch{Si} with different 
levels of generality.
A similar study is on-the-fly 
active learning~\cite{Vandermause2020On-the-FlyEvents}
which reached the RMSE of 
\SI{0.08}{\electronvolt\per\angstrom} with $133$ 
DFT calculations for \SI{10}{\pico\second} MD at 
\SI{620}{\kelvin}.
It is worth noting that this simulation based 
on normal GPR, using only two body and three body 
descriptors, took \SI{64}{\hour}.
Thus our SGPR algorithm is more than \SI{\sim300}{} 
times faster despite using the much more expensive 
SOAP descriptor and smaller computational resources 
(20 cores instead of 32).

RMSE of ML potentials usually increase with
temperature because of larger forces and
more disorder.
In \ch{Si} simulation above
from \SI{300}{} to \SI{700}{\kelvin}
RMSE increases from \SI{0.03}{} to 
\SI{0.07}{\electronvolt\per\angstrom}.
However, at higher temperatures larger RMSE
can be tolerated because of random forces 
due to thermal fluctuations.
In Ref~\cite{Behler2007} a silicon melt 
simulation with NN errors well above
\SI{0.2}{\electronvolt\per\angstrom}
accurately reproduced the DFT radial 
distribution function.
A minimal requirement is that distribution of 
forces error should be Gaussian centered at 0.
As such, they act as random forces due to 
thermal coupling with a heat bath 
and do not significantly alter the 
thermodynamics.

In order to show the performance of the algorithm in disordered phases,
we continued the simulation for \ch{Si} at \SI{1500}{} and \SI{2000}{\kelvin},
each for \SI{50}{\pico\second}.
At \SI{1500}{\kelvin}, the model did not change much 
(only 2 additional DFT calculations).
But at \SI{2000}{\kelvin}, the system melted which caused 
dozens of new DFT calculations.
After the melting completed, the model had learned the liquid phase
and only a few DFT calculations were invoked in the remaining steps (see \autoref{fig:si_melting}).
At the end, the model had accumulated 42 DFT data and 141 inducing LCEs,
and RMSE was \SI{\sim 0.15}{\electronvolt\per\angstrom} at \SI{2000}{\kelvin}.
In conclusion, the SGPR algorithm can learn the disordered
phases with a small number of DFT calculations as well.

\section{Sulphide solid electrolytes}

For an example of solid electrolytes,
a SGPR model is generated
by training in \ch{Li3PS4} ($\gamma$ and $\beta$ phases) 
and \ch{Li7P3S11} crystals.
During total of $\mathcal{O}(10^5)$ MD steps,
only \SI{111}{} FP data and 705 inducing LCEs 
are sampled.
To verify the model with data independent from training, 
we performed lengthy MLMD simulations for \ch{Li7P3S11}
at several temperatures in the range 
\SIrange[range-units=single]{300}{700}{\kelvin}
and \SI{e3}{} snapshots are randomly selected from 
trajectories.
For this testing set, RMSE=\SI{0.14}{\electronvolt\per\angstrom}
and $R^2=0.944$.
Training and testing are discussed in more
details in SM~\cite{prl_sm}.
The SGPR potential is perfectly stable, 
as demonstrated by the energy conservation test (see SM),
and during MD it never reached a non-physical 
arrangement of atoms which is a common worry 
for ML potentials.

\begin{figure}
    \centering
    \includegraphics[scale=0.9]{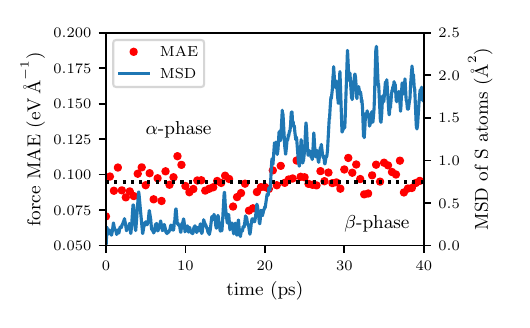}
    \caption{MAE of MLMD in the vicinity of $\alpha\to\beta$ phase 
    transition in \ch{Li7P3S11} at \SI{500}{\kelvin}
    ($2\times 1\times 1$ supercell).
    Since the transition occurs by rotation of \ch{S} atoms
    in \ch{P2S7} double-tetrahedra, MSD of \ch{S} atoms
    is a direct indicator.
    }
    \label{fig:alphabeta}
\end{figure}

We used this model for large scale
MLMD simulations of \ch{Li7P3S11} 
($4\times 2\times 2$ supercell, 672 atoms)
in isothermal-isobaric (NPT) ensemble
at several temperatures in the range
\SIrange[range-units=single]{300}{1200}{\kelvin}
and external pressure of \SI{e5}{\pascal}.
For this size, our implementation of MLMD
is $\mathcal{O}(10^4)$ faster than AIMD.
The mean squared displacement (MSD) of atoms,
defined by
\begin{equation}
    \mathrm{MSD}(t) = \frac{1}{N}\sum_{i=1}^{N}
    |\Delta\vec{r}_i(t)|^2
\end{equation}
where $N$ is the number of mobile ions and 
$\Delta\vec{r}_i(t)=\vec{r}_i(t)-\vec{r}_i(0)$,
is shown in SM.
A phase transition is detected by MLMD
(at $T\geq 450$) which occurs by 
rotations of the \ch{P2S7} double-tetrahedra
into a new orientational order.
In the following, the initial and the new structures
are referred to as $\alpha$ and $\beta$ phases.
The ground state energies of $\alpha$ and $\beta$ phases
obtained from exact DFT calculations (\texttt{VASP}/PBE)
are \SI{-4.413}{} and \SI{-4.416}{\electronvolt} per-atom
respectively which shows that they are almost iso-energetic
and that $\beta$ phase is not an artifact
of the SGPR potential.
Also, examples from $\beta$ phase are encountered
during training and testing.
\SI{0.14}{\electronvolt\per\angstrom}
is the lumped RMSE for both phases.
In \autoref{fig:alphabeta} the mean absolute error (MAE)
of ML is shown in the vicinity of $\alpha\to\beta$ transition.
The two steps in the MSD of \ch{S} atoms correspond
to delayed rotations of \ch{P2S7} double-tetrahedra in 
different layers.
The delay is \SI{\sim0.5}{\nano\second} at \SI{450}{\kelvin} 
with a $4\times 2\times 2$ supercell (see Fig. S7 in SM).
The $\alpha$- and $\beta$-\ch{Li7P3S11} 
crystal structure information files, 
visualizations, and radial distribution functions (RDFs)
are available as SM.
The $\alpha\to\beta$ transition is captured in a movie.
Specifically in the range
\SIrange[range-units=single]{450}{600}{\kelvin}
the system remains a few hundred \si{\pico\second} 
in the $\alpha$-phase before transition to the 
$\beta$-phase which allows us to estimate the 
\ch{Li} diffusivity in both phases.
In agreement with experimental reports~\cite{Mizuno2006HighLi2SP2S5, Yamane2007},
at \SI{900}{\kelvin}, the \ch{P} and \ch{S} atoms 
also start to diffuse, which indicates melting 
or its decomposition to more stable components
(see SM for RDF).

\begin{figure}
    \centering
    \includegraphics[scale=0.9]{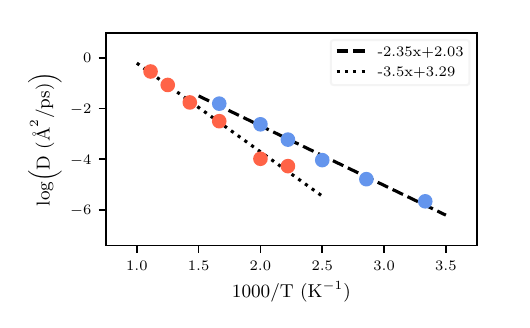}
    \caption{\ch{Li} diffusivity ($D$) in $\alpha$ (blue) 
    and $\beta$ (red) phases of \ch{Li7P3S11}.}
    \label{fig:ae}
\end{figure}

\begin{figure*}[t]
   \centering
    \begin{subfigure}[t]{0.23\textwidth}
        \centering
        \raisebox{0.5cm}{
        \includegraphics[width=\linewidth]{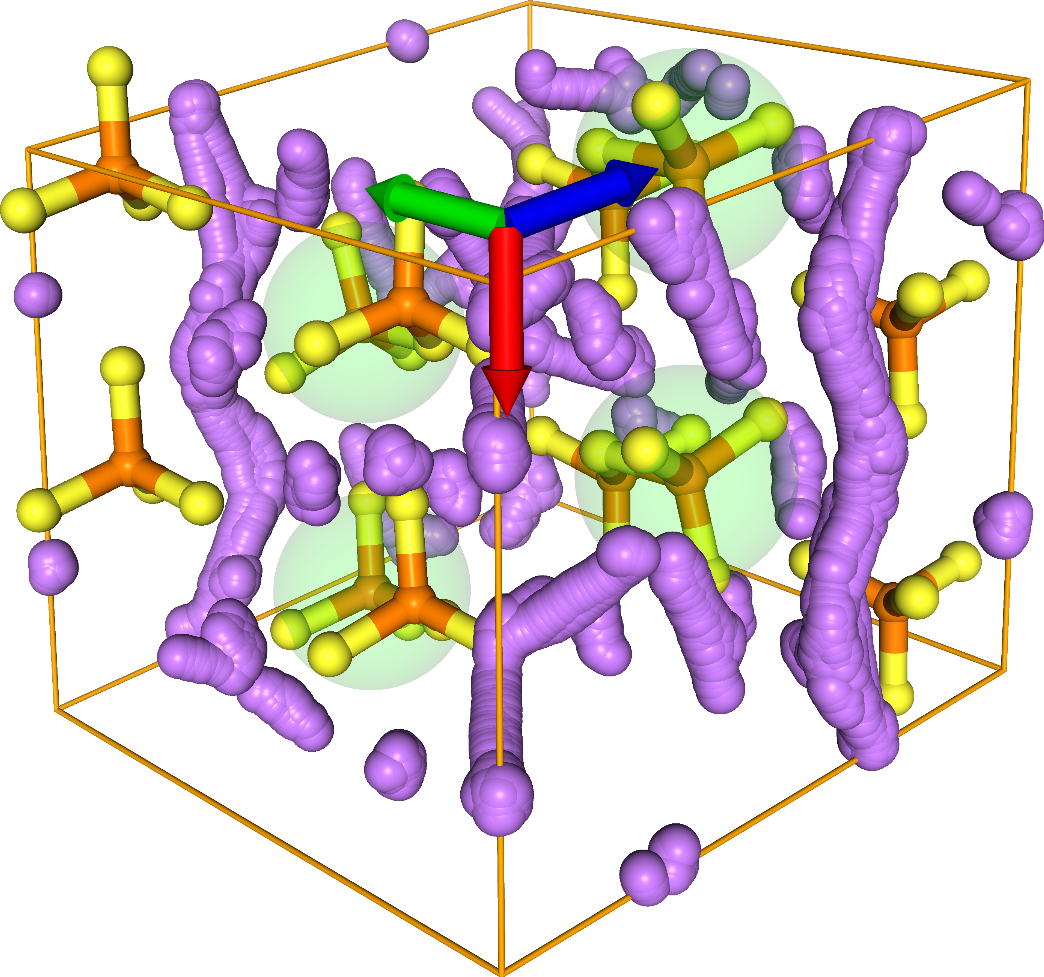}
        }
        \caption{} \label{fig:m_a}
    \end{subfigure}
    \hfill
    \begin{subfigure}[t]{0.23\textwidth}
        \centering
        \raisebox{0.5cm}{
        \includegraphics[width=\linewidth]{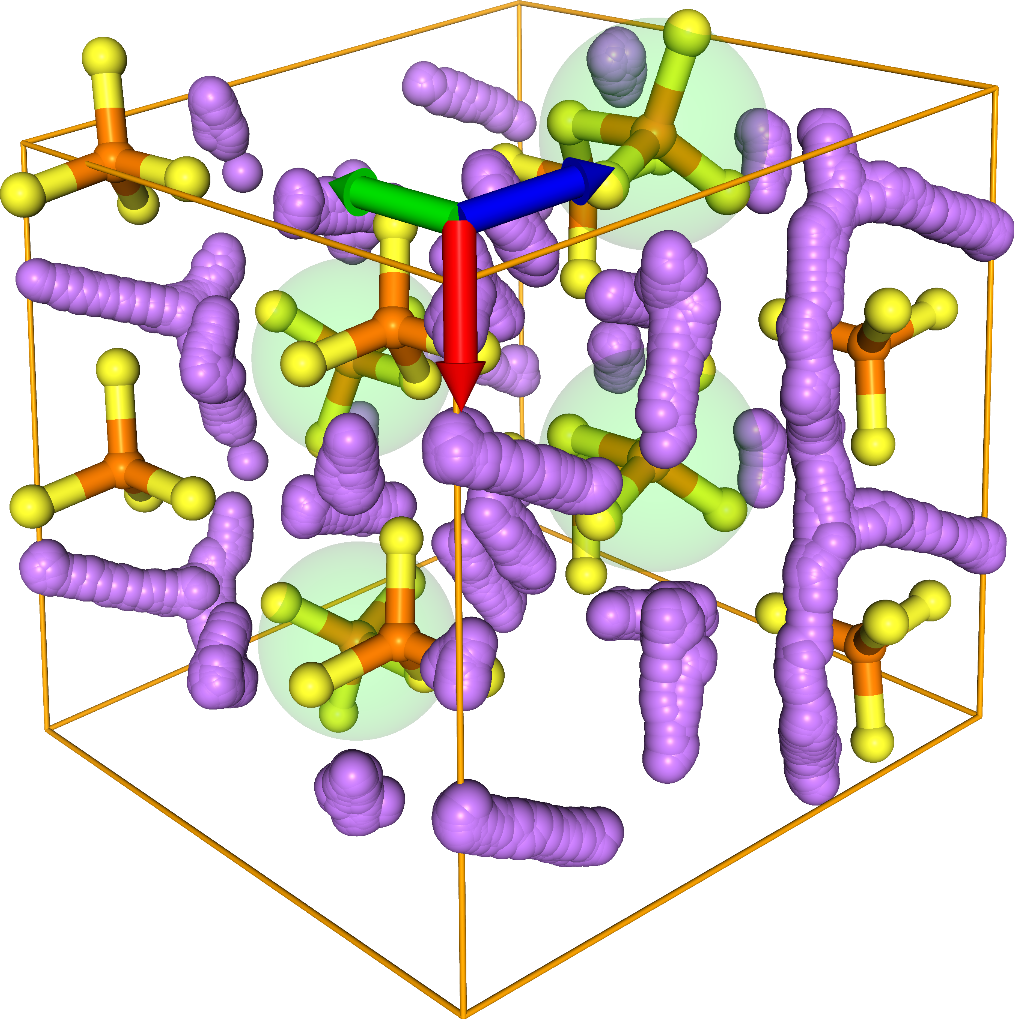}
        }
        \caption{} \label{fig:m_b}
    \end{subfigure}
    \hfill
    \begin{subfigure}[t]{0.40\textwidth}
        \centering
        \includegraphics[width=\linewidth]{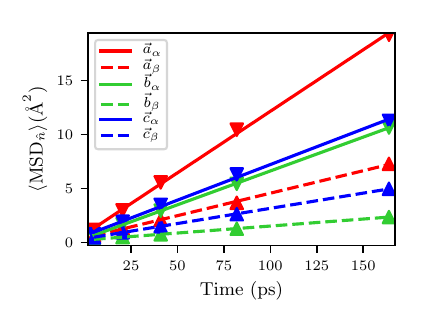}
        \caption{} \label{fig:abc}
    \end{subfigure}
    \caption{Time-lapse of the \ch{Li} positions over
    a few hundred \si{\pico\second} in 
    (a) $\alpha$ and (b) $\beta$ phases of \ch{Li7P3S11} with a
    $2\times1\times1$ cell at \SI{300}{\kelvin}.
    (c) Directional MSD
    in $\alpha$ and $\beta$ phases sampled by MD with a 
    $4\times2\times2$ cell at \SI{450}{\kelvin}.
    The respective unit cell vectors 
    $(\Vec{a}, \Vec{b}, \Vec{c})$
    are indicated with red, green and blue vectors
    in (a) and (b).}
    \label{fig:cells}
\end{figure*}

The diffusion coefficient (diffusivity) is defined as
\begin{equation} \label{eq:diffusion_coefficient}
    D = \lim_{t\to\infty} \frac{1}{2dt} 
    \big\langle\mathrm{MSD}(t)\big\rangle
\end{equation}
where $d=3$ is dimensions and 
$\langle\rangle$ indicates the ensemble average
(see {\autoref{fig:ae}}).
With the assumption of an Arrhenius temperature 
dependence for diffusivity
\begin{equation}
    D = D_0 e^{-E_a/k_BT} \texttt{,}
\end{equation}
we calculate the activation energies $E_a$ of 
\SI{0.20}{\electronvolt} and \SI{0.30}{\electronvolt} 
in $\alpha$ and $\beta$ phases, respectively.
The Nernst-Einstein relationship 
for conductivity $\sigma$ is given by
\begin{equation}\label{eq:NE}
    \sigma = \frac{Nq^2}{V k_BT} D
\end{equation}
where $N$ is number of \ch{Li} atoms, 
$V$  is volume, $q$ is ions electric charge,
and $T$ is temperature.
The ionic conductivity in $\alpha$ phase at 
RT(=\SI{300}{\kelvin}) becomes
\SI{3.5e-2}{\siemens\per\centi\meter} 
from direct simulation at this temperature
which is in reasonable agreement with 
previous reports of
\SI{5.7e-2}{}~\cite{Chu2016InsightsStudy} and 
\SI{4.5e-2}{\siemens\per\centi\meter}~\cite{Wang2017ComputationalSe}.
For $\beta$ phase, the RT conductivity 
is obtained by extrapolation of $D$ 
and approximating $V$ (which is $T$-dependent)
by the volume of the relaxed cell.
This yields the conductivity of 
\SI{2.3e-3}{\siemens\per\centi\meter} which is more 
than an order of magnitude lower than $\alpha$ phase.
Alternatively one can assume Arrhenius 
dependence for $\sigma T$ and extrapolate.
The numerical difference is $\mathcal{O}(1\%)$
since the volume of $\alpha$ phase expands
less than $1\%$ (\SI{300}{} to \SI{600}{\kelvin})
and $\beta$ phase about $\sim 4\%$
(\SI{450}{} to \SI{800}{\kelvin}).

Diffusion pathways are visualized in {\autoref{fig:cells}}.
In $\alpha$ phase, the diffusion of \ch{Li} atoms is highly 
concerted.
By reorientation of some of \ch{PS4} tetrahedra in $\beta$ phase,
this concerted diffusion is somewhat scattered.
For a more quantitative explanation,
the directional MSD along $\hat{n}$ is defined by
\begin{equation}
    \mathrm{MSD}_{\hat{n}}(t) = \frac{1}{N}\sum_{i=1}^{N}
    |\Delta\vec{r}_i(t).\hat{n}|^2
    \texttt{.}
\end{equation}
As shown in \autoref{fig:abc}, the diffusivity
is the largest along the $\vec{a}$-axis in both phases.
The main difference is that the diffusivity along $\vec{b}$ 
and $\vec{c}$ axes are almost equal in $\alpha$ phase,
while in $\beta$ phase this symmetry is clearly broken.
In Ref~\cite{Chu2016InsightsStudy}, it is pointed 
out that the one-dimensional MSD of \ch{Li} atoms 
along one of the axes in \ch{Li7P3S11} is nearly 
a factor of 2 higher than that along the other two axes.
This is consistent with the directional MSD that we 
report for $\alpha$-\ch{Li7P3S11} in \autoref{fig:abc}.
They also suggest that the \ch{Li} ionic motion is 
highly collective in agreement with our simulations.
The concerted motion of \ch{Li} atoms in 
$\alpha$ phase is also demonstrated in the
supplemented movie.

\begin{figure}
    \centering
    \includegraphics[scale=0.9]{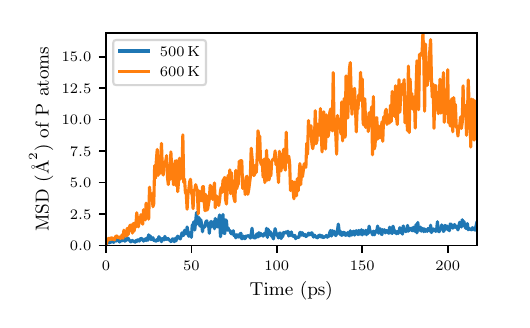}
    \caption{MSD of \ch{P} atoms in glass phase \ch{Li7P3S11}. 
    Diffusion of these atoms at \SI{600}{\kelvin}
    allows the glass to slowly relax and 
    eventually crystallize.}
    \label{fig:glas_cryst}
\end{figure}

The glass \ch{Li7P3S11} is created by a melt-quench 
simulation from \SI{1200}{} to \SI{300}{\kelvin}.
The \ch{Li} conductivity of 
\SI{9.3e-3}{\siemens\per\centi\meter}
is then obtained for the glass phase (with MLMD 
for \SI{3}{\nano\second}) which is higher than the 
conductivity in $\beta$ phase.
Equilibration of the glass at higher temperatures  
({\autoref{fig:glas_cryst}}) shows that
\ch{P} atoms start to diffuse at \SI{600}{\kelvin}
which indicates crystallization, in agreement with
experiment~\cite{Mizuno2006HighLi2SP2S5, Yamane2007, Seino2014}.
In liquid and glass phases 
RMSE=\SI{0.21}{\electronvolt\per\angstrom} 
and $R^2=0.917$ (see SM).

Experimentally, \ch{Li7P3S11} is often prepared in 
glass-ceramic phase by controlled crystallization
and heat treatment of the glass phase.
Depending on the heat treatment conditions, 
ionic conductivities of \SI{3.2e-3}{} and
\SI{1.7e-2}{\siemens\per\centi\meter}
are reported~\cite{Yamane2007, Seino2014}.
It was shown that the experimental conductivity
is correlated with the degree of crystallisation, 
depending on the heat treatment method~\cite{Seino2015AnalysisCeramic}.
Experimental measurements of the ionic conductivity
with techniques such as impedance spectroscopy 
probe the long-range dynamics of \ch{Li} which is 
influenced by the properties
such as grain boundaries, 
degree of amorphism, etc.
Wohlmuth et al.~\cite{Wohlmuth2015FastRelaxation}
probed the short-range diffusivity of \ch{Li}
in glass-ceramic \ch{Li7P3S11} 
and showed that the activation energy of 
bulk \ch{Li7P3S11} is \SI{0.20}{\electronvolt},
in agreement with our simulation.

\section{Conclusion}

In conclusion, we used the SGPR formalism
for generating accurate, stable, and scalable
interatomic potentials which, 
in the cases of \ch{Li7P3S11},
described the \ch{Li} diffusivity and
melting/crystallization (from glass phase)
temperatures, consistent with experiment.
An unchartered crystal structure is
found for this material which is nearly 
iso-energetic to the known crystal structure 
but has a much lower \ch{Li} diffusivity.
This phase should be avoided for better
\ch{Li} conductivity.
The SGPR algorithm is obtained by low-rank
approximation of the covariance matrix
in exact GPR and offers a much better 
scalability.
The computational cost of SGPR scales roughly 
similar to the Bayesian linear regression methods.
We also showed that converting BLR to SGPR is
straightforward and can be beneficial 
specially for acrive/on-the-fly machine learning
of interatomic potentials.

This work was supported by
NRF (National Honor Scientist Program: 2010-0020414),
UNIST (A.I. Incubation Project Fund: 1.210091.01),
and KISTI (KSSC-2018-CHA-0057, KSC-2019-CRE-0253, KSC-2020-CRE-0146).

%

\end{document}


\title{Supplemental Material for \\
``Towards Universal Sparse Gaussian Process Potentials: \\
Application to Lithium Diffusivity in 
Superionic Conducting Solid Electrolytes"}
\author{Amir Hajibabaei}
\author{Chang Woo Myung}
\author{Kwang S. Kim}
\email{kimks@unist.ac.kr}
\date{\today}

\maketitle

%

\section*{Including the forces data}
Let ${x}=\{\rho_i\}_{i=1}^{N}$ be the list of LCEs
for a given configuration $x$ and ${z}=\{\chi_j\}_{j=1}^{m}$
the list of inducing LCEs for SGPR.
The potential energy and forces obtained from
FP calculations are $E, f_i^\mu$
where $i$ indexes the atoms and $\mu$ the Cartesian
dimensions.
We define
\begin{equation}
    \beta(x, z) = \sum_{i=1}^N
    \begin{bmatrix}
    \kernel(\rho_i, \chi_1), &
    \ldots, &
    \kernel(\rho_i, \chi_m)
    \end{bmatrix}
    \texttt{.}
\end{equation}
Then the covariance and data matrices
for $x$ become
\begin{equation}\label{eq:knm_def}
    k_{xm} = 
    \begin{bmatrix}
    \beta(x,z) \\
    \vdots \\
    -\fpart{\beta(x, z)}{r_i^\mu} \\
    \vdots
    \end{bmatrix}
    ,
    Y_x = 
    \begin{bmatrix}
    {E} \\
    \vdots \\
    {f}_i^\mu \\
    \vdots
    \end{bmatrix}
\end{equation}
respectively, where $r_i^\mu$ are the coordinates of atoms.
These matrices are calculated for all configurations 
and are concatenated to build $k_{nm}$ and $Y$
defined in the main text.
\section*{The cost of kernel for GPR and SGPR}
For two configurations 
$A=\{a_i\}_{i=1}^{N_A}$ and $ B=\{b_i\}_{i=1}^{N_B}$
where $a_i, b_i$ are the per-atom descriptors,
covariance of total energies is calculated by
\begin{equation}
    \langle E_AE_B \rangle = \sum_{i,i'} \kernel(a_i,b_{i'}) 
    \to c \propto N_AN_B
    \texttt{,}
\end{equation}
where $c$ is the computational cost.
Let the average number of atoms in the neighborhood 
of each atom, within a cutoff radius $r_c$, be $n_c$.
Covariance of a total energy and a force component 
is calculated by
\begin{equation}
    \langle E_AF_{j'}^{\mu, B} \rangle = -
    \sum_{i,i'} \pdriv{\kernel(a_i,b_{i'})}{r_{j'}^\mu} 
    \to c\propto N_An_c
    \texttt{,}
\end{equation}
since the index $i'$ should cover all the atoms in 
the neighborhood of $j'$.
Similarly, covariance of two force components is 
calculated by
\begin{equation}
    \langle F_j^{\nu, A}F_{j'}^{\mu, B} \rangle = 
    \sum_{i,i'} \hess{\kernel(a_i,b_{i'})}{r_{j}^\nu}{r_{j'}^\mu} 
    \to c \propto n_c^2
    \texttt{,}
\end{equation}
The total cost for the evaluation of kernel between 
two configurations, considering that there are 
$3N_A$ and $3N_B$ force components, becomes
\begin{equation}
    \begin{split}
        c &\propto N_AN_B + (N_A n_c) \times (3N_B) + 
        (N_B n_c) \times (3N_A) + n_c^2 (3N_A)(3N_B) \\
        & \propto (1 + 3n_c)^2N_AN_B
    \end{split}
\end{equation}
In these equations the cost related to the length of
the descriptor vector is factored.

\textbf{GPR:}
In GPR we need to construct the full covariance 
matrix for the data given by 
${X}=\{{x}_i\}_{i=1}^{n}$.
The cost becomes
\begin{equation}
    c_1 \propto \sum_{i,j=1}^{n} (1 + 3n_c)^2N_iN_j
\end{equation}
If all of the configurations in the data have the same 
number of atoms $N$, the cost of constructing the 
full covariance matrix is proportional to
\begin{equation}
    c_1 \propto (1 + 3n_c)^2 N^2 n^2
\end{equation}

\textbf{SGPR:}
In SGPR, we only need to calculate the kernel between 
all of the configurations and a set of inducing local chemical 
environments (LCEs) ${z}=\{\rho_i\}_{i=1}^{m}$
as well as between LCEs themselves.
The cost is
\begin{equation}
    c_2 \propto (1 + 3n_c) N n m + m^2
\end{equation}

\textbf{Comparison:}
If we ignore $m^2$ in $c_2$
\begin{equation}
    \frac{c_2}{c_1} \propto \frac{m}{(1 + 3n_c) N n}
\end{equation}
For a typical choice of sizes $n=100$, $N=100$, $n_c=33$, 
and $m=1000$, which occur in our simulations,
the cost ratio becomes
\begin{equation}
    \frac{c_2}{c_1} = \frac{1}{1000}
\end{equation}

\section*{Adaptive Sampling Algorithm}
For adaptive sampling of the data ${X}$
and inducing LCEs ${z}$,
the model is built on-the-fly with MD
where the system $x(t)=\{\chi_j(t)\}_{j=1}^N$ 
evolves with the SGPR potential,
and FP calculations are carried out actively
to correct the model based on the following criterion.
The data and unique LCEs in the first step are 
automatically included in ${X}$ and ${z}$.
In the following steps, we try to insert on-the-fly 
LCEs to ${z}$.
A LCE $\chi_j$ is added to ${z}$ only 
if $\Delta E(\chi_j) > \epsilon$ where $\Delta E$
is the change in the predictive energy
resulting from inclusion of $\chi_j$ in ${z}$ 
and $\epsilon$ is a predetermined threshold.
Since it is too costly to try out all 
LCEs, in the spirit of importance sampling, 
we try insertion of LCEs based on their spilling factor 
(see main text) in a descending order and terminate 
at the first unsuccessful insertion trial.
If at least one LCE is added to ${z}$,
it means that the system has potentially crossed
into an unfamiliar region.
At this stage, in one algorithm (FAST), we calculated 
the exact FP energy and forces and 
included the snapshot $x(t)$ in ${X}$ 
only if $\Delta E(x(t)) > \epsilon$.
Thus, sometimes the exact FP data were regarded as 
redundant and were rejected by the model.
This was devised to keep the number of sampled data 
($\propto$ the computation cost in future steps) 
as small as possible.
In another algorithm (ULTRAFAST), in order to preempt 
unwanted FP calculations, predictions of the model were 
used as fake FP data and if the insertion 
was accepted they were corrected by exact calculations.
This algorithm is similar to
leave-one-out cross-validation method for active 
learning~\cite{Liu2018ADesign}.
For promoting global exploration and speeding up 
the sampling, we only sample at the extrema of 
the on-the-fly potential energy (see~\autoref{fig:fly}). 
Special care must be taken for stabilizing the model
in early steps.
\section*{Training and testing}
For true forces $\{f_i\}$ and the model's predictions $\{\widetilde{f}_i\}$,
coefficient of determination (CD) of the model is defined by
\begin{equation}
    R^2 = 1 - \frac{\sum_i (f_i-\widetilde{f}_i)^2}
    {\sum_i (f_i-\overline{f})^2} \texttt{.}
\end{equation}
Since CD is normalized by the variance in the data, 
it is a more consistent measure than 
root mean squared error (RMSE).
For energies, it is more appropriate to use mean absolute error
(MAE) because, unlike forces, energies are not distributed
around zero and uniform shift does not affect RMSE.

\begin{figure}
    \centering
    \includegraphics{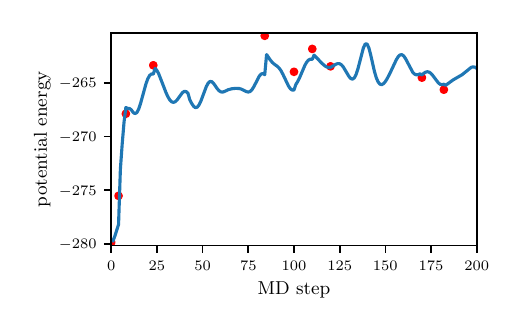}
    \caption{On-the-fly predicted potential energy of the SGPR model at
    the beginning of MD. 
    Red bullets indicate the timings and energies of 
    FP calculations.}
    \label{fig:fly}
\end{figure}

\begin{figure*}[h!]
    \centering
    \includegraphics[scale=0.8]{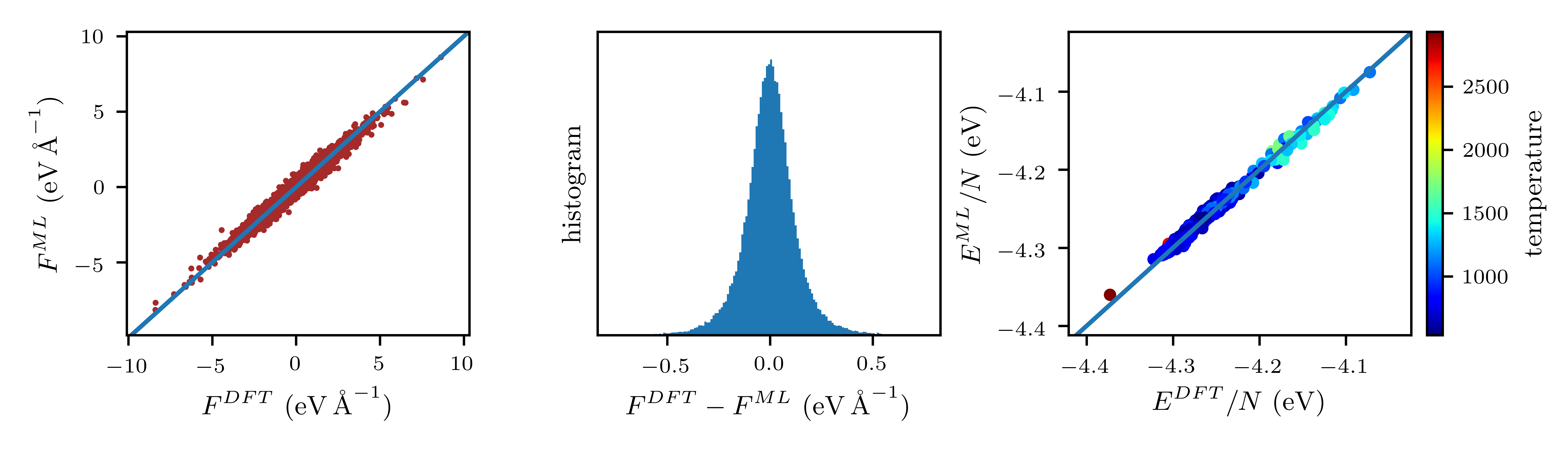}
    \caption{\textbf{Training data:} 
    testing the model trained with \ch{Li3PS4} 
    ($\alpha$ and $\beta$) and \ch{Li7P3S11} 
    on all 456 FP calculations which
    were performed during training. 
    For forces RMSE is \SI{0.13}{\electronvolt\per\angstrom} 
    and $R^2=0.967$.
    For energies MAE is \SI{3.817e-3}{\electronvolt} per-atom.
    }
    \label{fig:train}
\end{figure*}

As the first example in the main text,
a SGPR potential is trained on-the-fly with MD,
with three different systems:
$\gamma$-\ch{Li3PS4} with $(1 \times 1 \times 2)$,
$\beta$-\ch{Li3PS4} with $(2 \times 1 \times 1)$,
and \ch{Li7P3S11} with $(2 \times 1 \times 1)$ cells.
For sampling, the systems were quenched from
$\sim$\SI{1500}{\kelvin} to \SI{700}{\kelvin} 
in \SI{\sim 5}{\pico\second}
by velocity re-scaling 
and kept at this temperature
for \SI{\sim 45}{\pico\second} using a time step of 
\SI{2}{\femto\second}.
In quenching, there is no time for the crystal
to melt (since atoms do not have time to displace too far
the crystal becomes only locally disordered)
but this local distortion results in stream of diverse
LCEs to the sampling algorithm which helps in
faster stabilization of the model at the beginning 
of training.
This practice is beneficial but not compulsory.
The algorithm FAST (see adaptive sampling algorithms 
in the main text) is applied in training (see~\autoref{fig:fly}), 
therefore in total only \SI{456} FP calculations 
are performed from which only \SI{111}{} are 
sampled by the model (as data $\mathrm{X}, Y$) and
705 LCEs are sampled as the inducing set ($\mathrm{z}$).
For the sake of speed,
only $\Gamma$ point is selected for sampling k-space,
but afterwards the potential energy and forces are 
recalculated with a $2\times 2\times 2$ k-point grid 
only for those \SI{111}{} data 
which are included in the model.
This corrects the forces by an RMSE of about
\SI{0.04}{\electronvolt\per\angstrom}.
In \Autoref{fig:train} performance of the 
model on all \SI{456} examples is demonstrated.

\begin{figure*}[h!]
    \centering
    \includegraphics[scale=0.8]{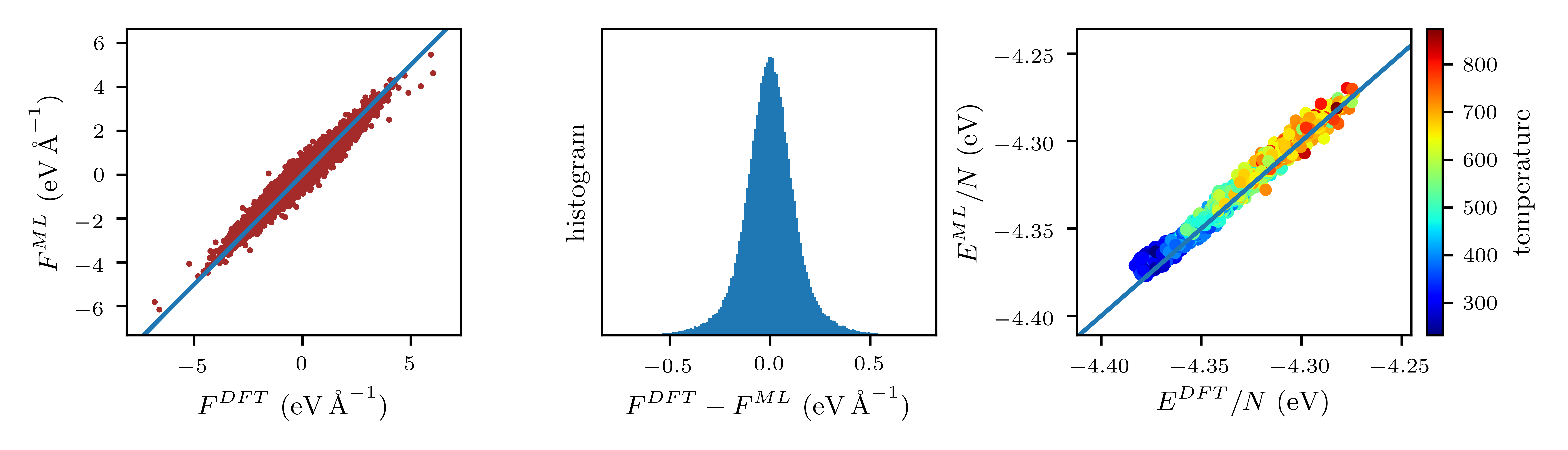}
    \caption{\textbf{Crystal phase:} 
    \SI{1000}{} samples are randomly selected from 
    extensive MLMD simulations of \ch{Li7P3S11} 
    at five temperatures in the 
    range \SIrange[range-units=single]{300}{700}{\kelvin}
    for \SIrange[range-units=single]{0.8}{2.2}{\nano\second}.
    The number of atoms in these samples is 84.
    For forces RMSE is \SI{0.139}{\electronvolt\per\angstrom} 
    and $R^2=0.944$.
    For energies MAE is \SI{3.822e-3}{\electronvolt} per-atom.
    }
    \label{fig:test}
\end{figure*}

To verify the model with independent data, 
MLMD simulations at five temperatures in the 
range \SIrange[range-units=single]{300}{700}{\kelvin}
for \SIrange[range-units=single]{0.8}{2.2}{\nano\second}
are performed (only for \ch{Li7P3S11}, same size as training) 
and \SI{e3}{} snapshots are randomly selected.
Accuracy of the model is tested with these samples 
in \Autoref{fig:test}.

\begin{figure*}[h!]
    \centering
    \includegraphics[scale=0.8]{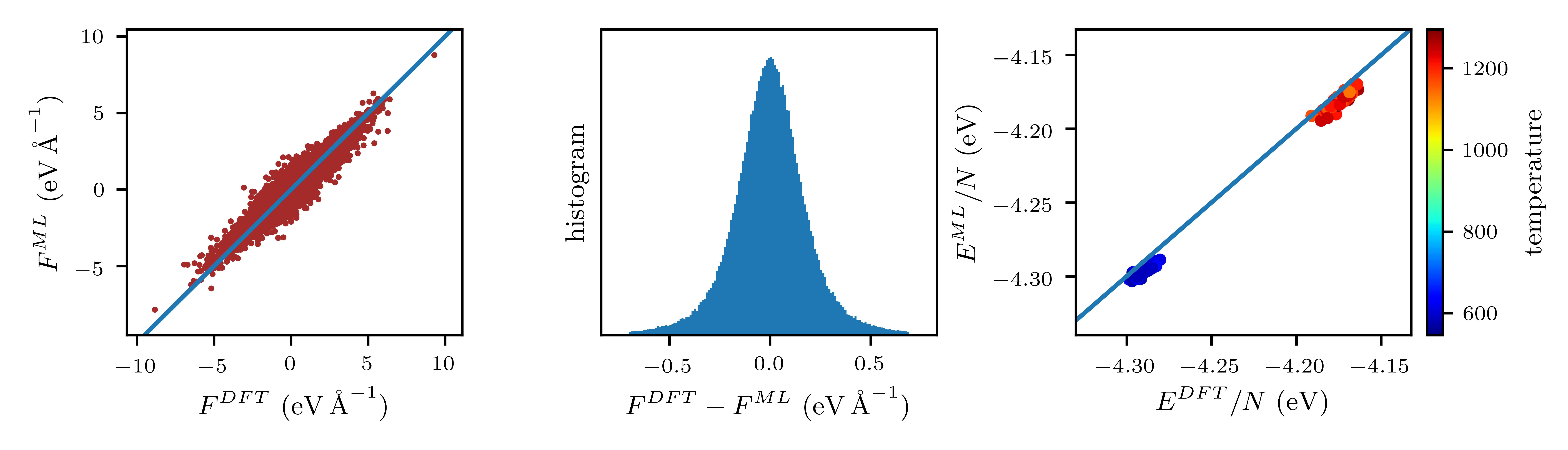}
    \caption{\textbf{Liquid and glass phases:}
    50+70 samples are drawn from MLMD simulation 
    of the liquid phase \ch{Li7P3S11} at \SI{1200}{\kelvin} and
    glass phase (obtained after melt-quenching)
    at \SI{600}{\kelvin} for \SI{\sim 0.2}{\nano\second}.
    The number of atoms in these samples is 672.
    The errors in these phases are similar.
    For forces RMSE is \SI{0.21}{\electronvolt\per\angstrom}
    and $R^2=0.917$.
    For energies MAE is \SI{6.35e-3}{\electronvolt} per-atom.
    }
    \label{fig:glass}
\end{figure*}

In addition, the liquid and glass phase \ch{Li7P3S11}
are simulated with the SGPR model.
Testing of the model in these phases is shown 
in \Autoref{fig:glass}.
Since size of the cells in these simulations (672 atoms)
are larger than the training systems, the SGPR model
slightly underestimates the total energy 
(by \SI{\sim 3e-3}{\electronvolt} per-atom)
due to the long-range forces that are excluded
because of a finite cutoff radius.
Treatment of the long-range forces with ML models
is not trivial yet.

\begin{figure*}[h!]
    \centering
    \includegraphics[width=\linewidth]{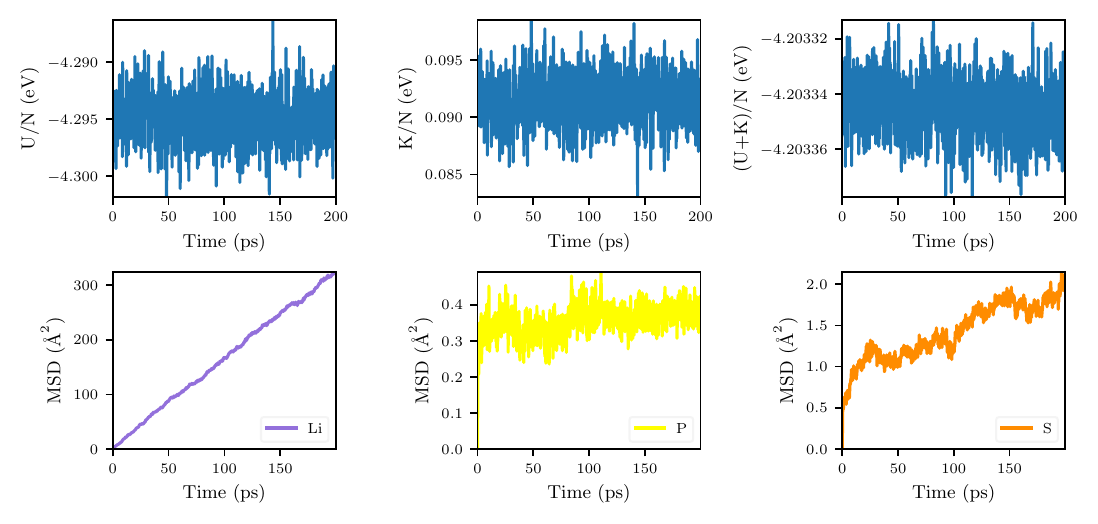}
    \caption{\textbf{Energy conservation:}
    MLMD simulation of \ch{Li7P3S11} with 672 atoms 
    in NVE ensemble with velocity-Verlet algorithm. 
    The temperature is \SI{\sim 700}{\kelvin}.
    The per-atoms potential energy, kinetic energy, and total
    energy are plotted in the first row.
    The total energy is conserved up to
    small fluctuations of \SI{\sim e-5}{\electronvolt} per atom
    which are related to the finite timestep of \SI{2}{\femto\second}.
    In the second row, MSDs of atoms from their 
    initial positions are plotted.
    }
    \label{fig:nve}
\end{figure*}

\begin{figure*}[h!]
    \centering
    \begin{subfigure}[t]{0.19\textwidth}
        \centering
        \raisebox{1.0cm}{
        \includegraphics[width=\linewidth, angle=90]{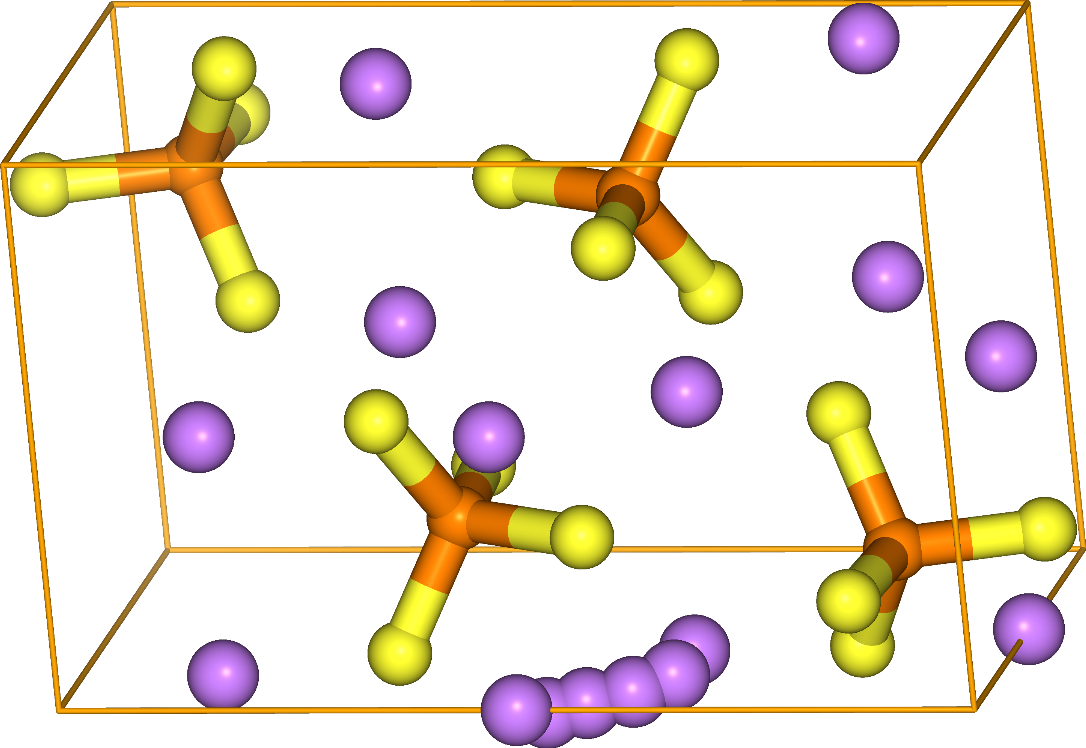}
        }
    \end{subfigure}
    \begin{subfigure}[t]{0.3\textwidth}
        \centering
        \includegraphics[width=\linewidth]{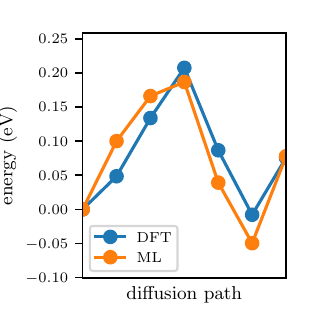}
    \end{subfigure}   
    \caption{
    Nudged elastic band calculations for \ch{Li} hopping
    to a neighboring vacancy (created by \ch{Li} removal)
    in $\gamma$-\ch{Li3PS4} with DFT and ML potentials.
    $(2\times2\times2)$ k-point grid is applied in DFT calculations.}
    \label{fig:neb}
\end{figure*}

To demonstrate that the generated SGPR model is 
energy-conservative, an MD simulation in NVE
ensemble is carried out which is shown in 
\Autoref{fig:nve}. 
A nudged elastic band (NEB) calculations is 
also carried out which is directly compared 
with DFT in \Autoref{fig:neb}.

\FloatBarrier
\section*{Large scale MLMD simulations}
\begin{figure*}[h!]
    \centering
    \includegraphics[width=\linewidth]{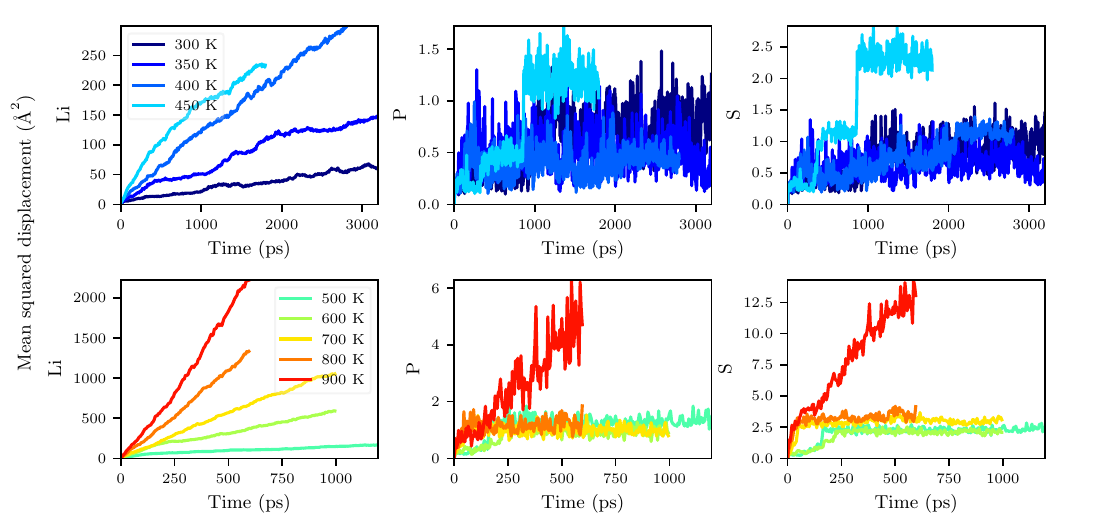}
    \caption{Mean squared displacements of atoms 
    in \ch{Li7P3S11} system (with 672 atoms)
    during the MD simulation in the NPT ensemble 
    with temperatures ranging from
    \SIrange[range-units=single]{300}{900}{\kelvin}.
    The jumps in the MSD of \ch{P} and \ch{S} atoms are
    associated with a crystal-crystal ($\alpha\to\beta$)
    phase transition.
    The crystal melts at \SI{900}{\kelvin}.}
    \label{fig:msd}
\end{figure*}

\begin{figure}
    \centering
    \includegraphics[width=\textwidth]{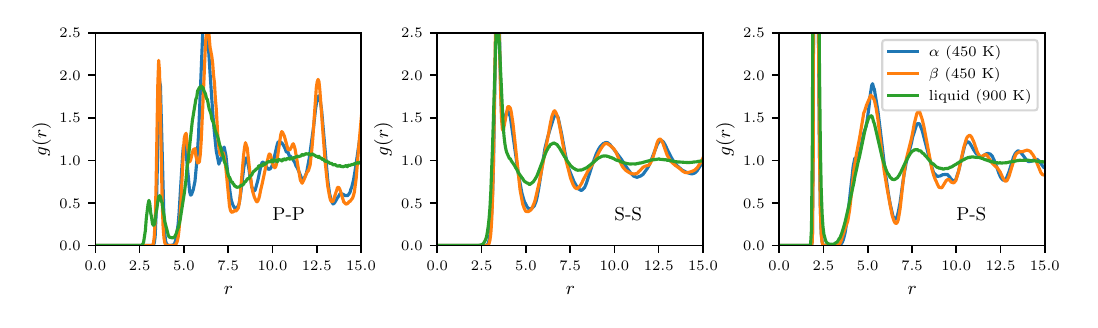}
    \caption{Radial distribution functions in 
    $\alpha$, $\beta$, and molten phases of \ch{Li7P3S11}.}
    \label{fig:rdf}
\end{figure}

\begin{figure*}[h]
   \centering
    \begin{subfigure}[t]{0.3\textwidth}
        \centering
        \includegraphics[width=\linewidth]{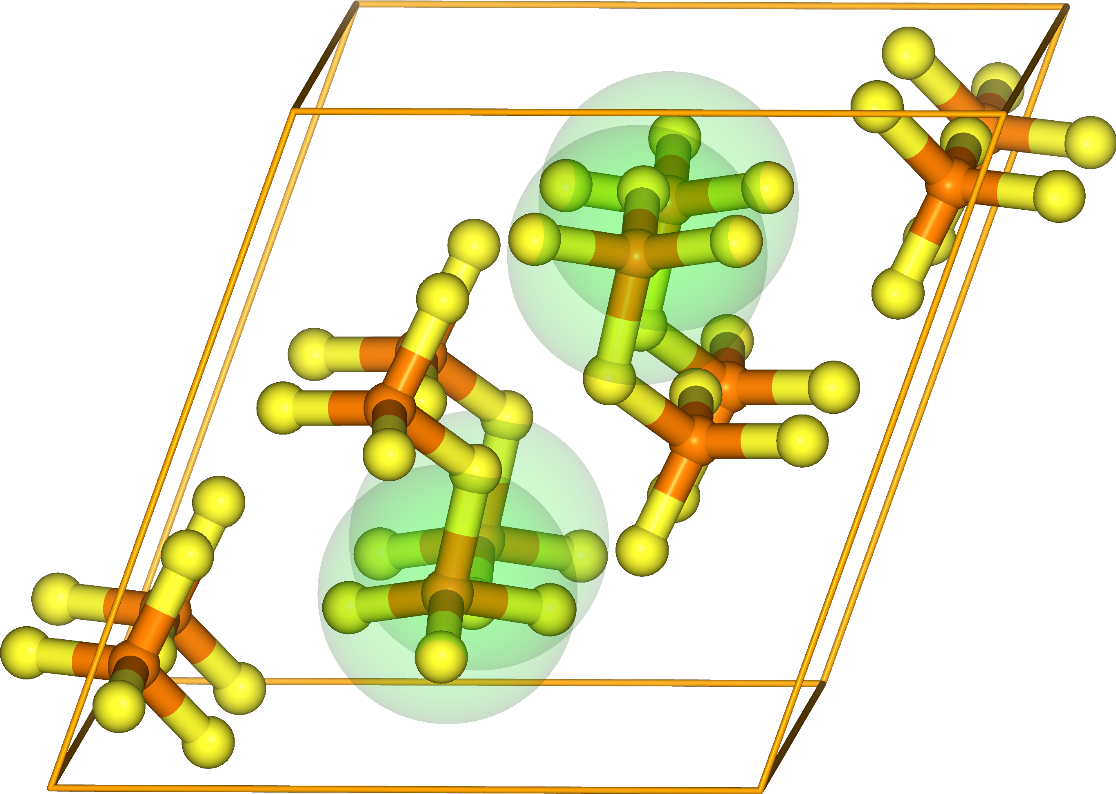}
        \caption{} \label{fig:cell_a}
    \end{subfigure}
    \begin{subfigure}[t]{0.3\textwidth}
        \centering
        \includegraphics[width=\linewidth]{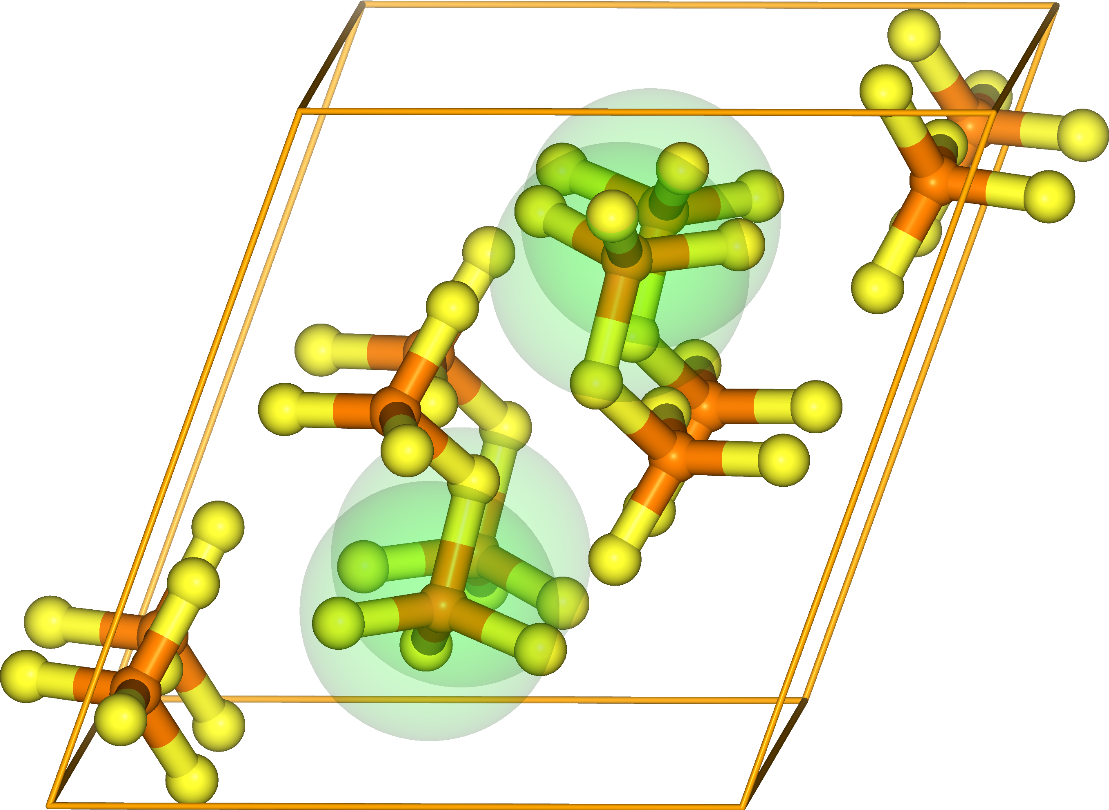}
        \caption{} \label{fig:cell_b}
    \end{subfigure}
    \caption{Relaxed (a) $\alpha$ and (b) $\beta$ cells
    of \ch{Li7P3S11}, 
    while \ch{Li} atoms are removed for clarity.
    The rotated \ch{PS4} tetrahedra are highlighted with green.}
    \label{fig:cells}
\end{figure*}

\begin{figure}[h!]
    \centering
    \includegraphics[width=5cm]{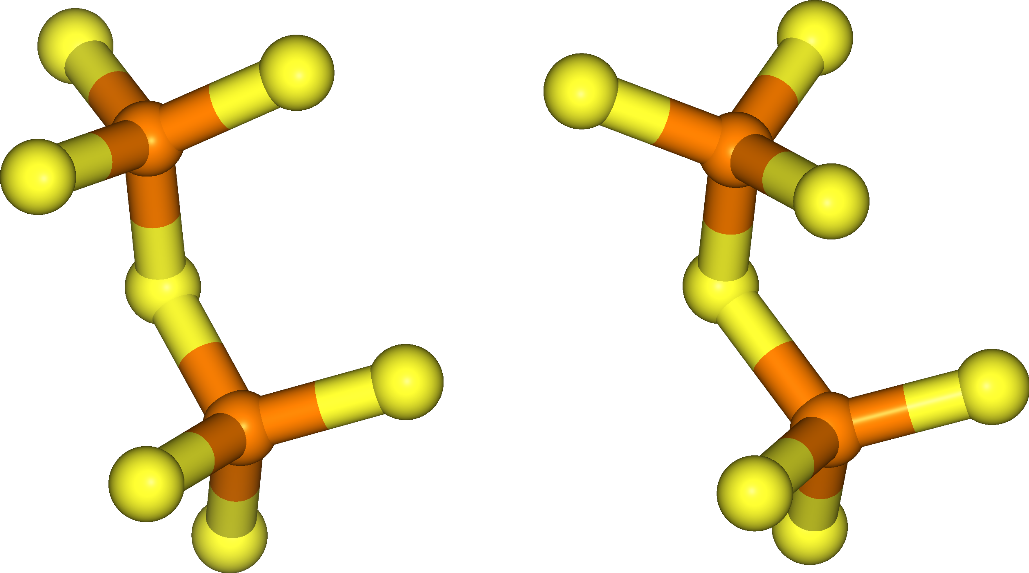}
    \caption{Preferred orientations of \ch{P2S7}
    double-tetrahedra in $\alpha$ (left) and 
    $\beta$ (right) phases of \ch{Li7P3S11}.
    }
    \label{fig:p2s7}
\end{figure}

\FloatBarrier

\section*{Computational Performance}
For a system with $N$ atoms, the computational 
cost of predicting energy and forces with the 
SGPR model is $\mathcal{O}(N)$,
while the complexity of FP calculations is 
$\mathcal{O}(N^3)$ or at best 
$\mathcal{O}(N^2\log N)$.
The elapsed time for FP calculations of the
energy and forces for a system with 672 atoms 
was \SI{9718}{\second} using VASP with 80 cores,
while the elapsed time for the same calculation with 
SGPR was \SI{5.3}{\second} using only 20 cores.
The cost defined by ${\mathrm{elapsed}\times\mathrm{cores}}/{N}$
is $\sim \SI{0.15}{\second}$ for SGPR which is 
better than FP calculations by a factor of $\mathcal{O}(10^4)$.

\FloatBarrier
%